\documentclass[12pt]{article}

\usepackage{amsmath,amssymb,amsthm}

\newcommand{\vr}{\boldsymbol{r}}
\newcommand{\vj}{\boldsymbol{j}}
\newcommand{\vv}{\boldsymbol{v}}
\newcommand{\vp}{\boldsymbol{p}}
\newcommand{\vomega}{\boldsymbol{\omega}}
\newcommand{\vrho}{\boldsymbol{\rho}}
\newcommand{\Phiv}{\varPhi}
\newcommand{\Psiv}{\varPsi}
\newcommand{\Gammav}{\varGamma}

\newtheorem{thm}{\sc Theorem\!\!}

\begin{document}

\title{\bf ``Velocities'' in Quantum Mechanics}

\author{Toshiki Shimbori$^*$ and Tsunehiro Kobayashi$^\dag$ \\
{\footnotesize\it $^*$Institute of Physics, University of Tsukuba}\\
{\footnotesize\it Ibaraki 305-8571, Japan}\\
{\footnotesize\it $^\dag$Department of General Education 
for the Hearing Impaired,}
{\footnotesize\it Tsukuba College of Technology}\\
{\footnotesize\it Ibaraki 305-0005, Japan}}

\date{}

\maketitle

\begin{abstract}
 The present paper deals with some kind of quantum ``velocity'' 
 which is introduced by the method of hydrodynamical analogy. 
 It is found that this ``velocity'' is in general irrotational, 
 namely, a vorticity vanishes, 
 and then a velocity potential must exist in quantum mechanics. 
 In some elementary examples of stable systems 
 we will see what the ``velocities'' are. 
 In particular, the two-dimensional flows of these examples 
 can be expressed by complex velocity potentials 
 whose real and imaginary parts are 
 the velocity potentials and stream functions, respectively. 
\end{abstract}

\thispagestyle{empty}

\setcounter{page}{0}

\pagebreak

 It is well known that the probability density $\rho(t,\vr)$ and 
 the probability current $\vj(t,\vr)$ of a state $\psi(t,\vr)$ 
 in non-relativistic quantum mechanics 
 are given by~\cite{dirac,landau,jjs}
 \begin{align}
  \rho(t,\vr)&\equiv\left| \psi(t,\vr)\right|^2, \label{3.1.1}\\
  \vj(t,\vr)&\equiv\left.\Re\left[\psi(t,\vr)^*
  \left(-i\hslash\nabla\right)\psi(t,\vr)\right]\right/m, 
  \label{3.1.2}
 \end{align}
 where $m$ is the mass of the particle. 
 They satisfy the equation of continuity 
 \begin{equation}
  \frac{\partial\rho}{\partial t}+\nabla\cdot\vj=0. 
   \label{3.1.3}
 \end{equation}
 However, these $\rho$ and $\vj$ are independent of the time $t$ 
 when considering state is a closed state of the stable system 
 (e.g. the harmonic oscillator, the hydrogen atom) or 
 a stationary state (e.g. the free particle). 
 For such closed stationary states 
 the probability currents will be at rest or uniform, 
 and the equation \eqref{3.1.3} reduces to 
 $$\nabla\cdot\vj=0. $$ 
 Thus the probability currents are simple 
 for the closed stationary states of the stable system. 
 The present paper deals with some kind of ``{\it velocity}'', 
 which is different from the eigenvalue of the velocity operator and 
 which is a stepping-stone to a more description 
 in quantum mechanics. 
 
 We start out with the equations of 
 hydrodynamics, consisting of 
 Euler's equation of continuity for the density and velocity of fluid 
 and so on. 
 Let us try to introduce a ``velocity'' 
 which will be the analogue of the hydrodynamical one. 
 In hydrodynamics~\cite{lamb,landau2,batchelor}, the fluid at one time 
 can be represented by 
 the density $\rho$ and the fluid velocity $\vv$. 
 They satisfy Euler's equation of continuity 
 \begin{equation}
  \frac{\partial\rho}{\partial t}+\nabla\cdot(\rho\vv)=0. 
   \label{3.1.4}
 \end{equation} 
 Comparing this equation with \eqref{3.1.3}, 
 we are thus led to the following 
 {\it definition for the quantum} ``{\it velocity}'' 
 {\it of a state} $\psi(t,\vr)$, 
 \begin{equation}
  \text{``$\vv$''}\equiv\frac{\vj(t,\vr)}{\left| \psi(t,\vr)\right|^2}, 
   \label{3.1.5}
 \end{equation}
 in which $\vj(t,\vr)$ is given by \eqref{3.1.2}. 
 It is stressed that this ``velocity'' is different from 
 the eigenvalue of the velocity operator $\hat{\vv}=-i\hslash\nabla/m$, 
 except for the cases that $\psi(t,\vr)$ is an eigenfunction of 
 the momentum operator $\hat{\vp}=-i\hslash\nabla$. 
 Note that, if we can separate variables of $\psi(t,\vr)$, 
 then ``$\vv$'' does not contain the time $t$ explicitly. 
 
 Equation \eqref{3.1.5} is justified from the following point of view. 
 In semiclassical cases 
 the time-dependent wave function can be written~\cite{dirac,landau,jjs} 
 \begin{equation}
  \psi(t,\vr)=\sqrt{\rho}\, e^{iS/\hslash}, \label{3.1.6} 
 \end{equation}
 where $\rho$ is the probability density \eqref{3.1.1} and 
 $S$ is the quantum analogue of the classical action 
 which is a real function of the $t$ and $\vr$. 
 The ``velocity'' \eqref{3.1.5} gives 
 \begin{equation}
  \text{``$\vv$''}=\nabla S/m. \label{3.1.7} 
 \end{equation}
 The right-hand side is known just as the velocity 
 in classical dynamics. 
 Sakurai~\cite{jjs} introduced the ``velocity'' 
 by this relation \eqref{3.1.7}. 
 
  We shall now consider the {\it vorticity} in quantum mechanics. 
  It is defined as in the hydrodynamics, by 
  \begin{equation}
   \vomega\equiv\nabla\times\text{``$\vv$''}. \label{3.1.8}
  \end{equation} 
  With the above definition we have the next theorem. 
  
  \begin{thm}
   If the variables $t$, $\vr$ of a time-dependent wave function 
   $\psi(t,\vr)$ can be separated, then 
   \begin{equation}
    \vomega=0 \label{3.1.t}
   \end{equation}
   for the domain in which the ``velocity'' 
   does not have singularities. 
  \end{thm}
  
  To prove the theorem, 
  it is sufficient to evaluate $\omega$ 
  in the two-dimensional $xy$ space. 
  From the hypothesis, the time-dependent wave function 
  must be of the form 
  $$\psi(t,x,y)=T(t) X(x) Y(y). $$
  The components of \eqref{3.1.2} now give 
  \begin{align*}
   j_x(t,x,y)&=\left|T(t)\right|^2 \left|Y(y)\right|^2 
   \Re\left.\left[ X(x)^* 
   (-i\hslash) X(x)^\prime\right]\right/m, \\
   j_y(t,x,y)&=\left|T(t)\right|^2 \left|X(x)\right|^2 
   \Re\left.\left[ Y(y)^* 
   (-i\hslash) Y(y)^\prime\right]\right/m. 
  \end{align*}
  The components of \eqref{3.1.5} then give 
  \begin{align*}
   \text{``$v_x$''}&=\Re\left[ X(x)^* 
   (-i\hslash) X(x)^\prime\right]
   \left/m\left|X(x)\right|^2\right. , \\
   \text{``$v_y$''}&=\Re\left[ Y(y)^* 
   (-i\hslash) Y(y)^\prime\right]
   \left/m\left|Y(y)\right|^2\right. , 
  \end{align*} 
  showing that each component of ``velocity'' 
  is depending only on the corresponding Cartesian coordinate. 
  When this ``velocity'' is not singular, 
  the result of its rotation is 
  $$\omega=\frac{\partial\text{``$v_y$''}}{\partial x}
  -\frac{\partial\text{``$v_x$''}}{\partial y}=0. $$
  Thus the theorem is proved in two-dimensional space. 
  
  Theorem is still valid in three-dimensional space 
  or in terms of polar coordinates, 
  only formal changes being needed in the proof. 
  
  In hydrodynamics~\cite{lamb,landau2,batchelor}, 
  the flow satisfying $\vomega=0$ 
  is called {\it potential flow} or {\it irrotational flow}. 
  Thus this theorem asserts that {\it the flow in quantum mechanics 
  is in general irrotational flow, 
  when all variables are separable}. 
  The ``velocity'' in irrotational flow satisfying \eqref{3.1.t} 
  may be described by the gradient of the 
  {\it velocity potential} $\Phiv$, 
  \begin{equation}
   \text{``$\vv$''}=\nabla\Phiv. \label{3.1.9}
  \end{equation}
  Comparing this with \eqref{3.1.7}, we see that 
  \begin{equation}
   \Phiv =S/m, \label{3.1.10} 
  \end{equation}
  in semiclassical cases. 
  The right-hand side here is undefined to the extent of 
  an arbitrary additive real constant. 
  
  We now proceed to study only the {\it two-dimensional} or 
  {\it plane flow}. 
  Let us consider the ``velocity'' \eqref{3.1.5} 
  which is solenoidal, namely 
  \begin{equation}
   \nabla\cdot\text{``$\vv$''}\equiv 
    \frac{\partial\text{``$v_x$''}}{\partial x} 
    +\frac{\partial\text{``$v_y$''}}{\partial y}=0. \label{3.1.11} 
  \end{equation}
  The ``velocity'' in two-dimensional flow satisfying \eqref{3.1.11} 
  may be described by the rotation of the 
  {\it stream function} $\Psiv$, 
  \begin{equation}
   \text{``$v_x$''}=\frac{\partial\Psiv}{\partial y}, \,\,\, 
    \text{``$v_y$''}=-\frac{\partial\Psiv}{\partial x}. \label{3.1.12}
  \end{equation}
  
  Further, in the two-dimensional irrotational flow, 
  by combining \eqref{3.1.12} with \eqref{3.1.9} we get 
  \begin{align*}
   \text{``$v_x$''}&=\frac{\partial\Phiv}{\partial x}
   =\frac{\partial\Psiv}{\partial y}, \\
   \text{``$v_y$''}&=\frac{\partial\Phiv}{\partial y}
   =-\frac{\partial\Psiv}{\partial x}. 
  \end{align*}
  It is known as {\it Cauchy-Riemann's equations} 
  between the velocity potential and the stream function. 
  We can therefore take the 
  {\it complex velocity potential} 
  \begin{equation}
   W(z)=\Phiv(x,y)+i\Psiv(x,y), \label{3.1.13}
  \end{equation}
  which is a regular function of the complex variable $z=x+iy$. 
  The differentiation of $W(z)$ gives us the {\it complex velocity} 
  \begin{equation}
   \frac{dW}{dz}=\text{``$v_x$''}-i\text{``$v_y$''}. \label{3.1.14}
  \end{equation}
  In this way we know that 
  in the two-dimensional irrotational flow 
  it is advantageous to use the theory of functions of 
  a complex variable~\cite{lamb,landau2,batchelor}. 
  
  The solenoidal condition \eqref{3.1.11} holds 
  for incompressible fluids in hydrodynamics, 
  since $\rho$ is a constant. 
  In quantum mechanics, however, 
  the probability density \eqref{3.1.1} depends generally on $\vr$, 
  so that the ``velocity'' \eqref{3.1.5} is not always solenoidal. 
  
  We shall illustrate the physical value of ``velocities'' 
  by applying them to examine 
  some elementary examples of stable systems. 
  
    \paragraph{Example 1. The free particle \\}
    Let us first consider the free particle in two dimensions 
    as an example of a stationary state. 
    The plane wave is of the form 
    \begin{equation}
     u_{p_x p_y}(x,y)=a e^{i(p_x x+p_y y)/\hslash}, 
      \label{3.e.1} 
    \end{equation}
    where $a$ is independent of $t$, $x$ and $y$. 
    The probability current of the plane wave is 
    $$j_x(x,y)=|a|^2 p_x/m, \,\,\, j_y(x,y)=|a|^2 p_y/m, $$
    and hence their ``velocity'' is 
    $$\text{``$v_x$''}=p_x/m, \,\,\, \text{``$v_y$''}=p_y/m. $$
    Equations \eqref{3.1.t} and \eqref{3.1.11} 
    are easily seen to hold for the plane wave. 
    Therefore the velocity potential satisfying \eqref{3.1.9} 
    or \eqref{3.1.10} of the plane wave is 
    \begin{equation}
     \Phiv=(p_x x+p_y y)/m, \label{3.e.4} 
    \end{equation}
    and the stream function satisfying \eqref{3.1.12} is 
    \begin{equation}
     \Psiv=(p_x y-p_y x)/m. \label{3.e.5} 
    \end{equation}
    For the state represented by \eqref{3.e.1}, 
    the complex velocity potential \eqref{3.1.13} gives, 
    from \eqref{3.e.4} and \eqref{3.e.5} 
    \begin{align}
     W&=(p_x x+p_y y)/m +i(p_x y-p_y x)/m \notag \\
     &=(p_x -ip_y)z/m. \label{3.e.6}
    \end{align}
    According to  hydrodynamics~\cite{lamb,landau2,batchelor}, 
    the flow round the angle $\pi/n$ 
    is expressed by the complex velocity potential 
    \begin{equation}
     W=A z^n, \label{3.e.7} 
    \end{equation}
    $A$ being a number. 
    Equation \eqref{3.e.6} is of the form \eqref{3.e.7} with $n=1$, 
    and it shows that {\it the complex velocity potential 
    of the plane wave just expresses uniform flow}. 
    
    \paragraph{Example 2. The harmonic oscillator \\}
    We shall now consider the eigenstate of 
    the two-dimensional harmonic oscillator 
    as an example of a closed state of the stable system. 
    The eigenfunction is,\footnote{The $\omega$ here, denoting 
    the angular frequency, is, of course, to be distinguished 
    from the $\omega$ denoting the vorticity.} in terms of 
    the Cartesian coordinates $x$, $y$, 
    \begin{equation}
     u_{n_x n_y}(x,y)=N_{n_x} N_{n_y} e^{-\alpha^2(x^2 +y^2)/2} 
      H_{n_x}(\alpha x) H_{n_y}(\alpha y)  \,\,\, 
      \left(\alpha\equiv\sqrt{m\omega/\hslash}\right),  
      \label{3.e.8} 
    \end{equation}
    where $N_{n_x}$, $N_{n_y}$ are the normalizing constants. 
    But the Hermite polynomials $H_{n_x}(\alpha x)$, $H_{n_y}(\alpha y)$
    are real functions of $x$, $y$, respectively. 
    Thus the probability current of the harmonic oscillator vanishes 
    and their ``velocity'' also vanishes. 
    Therefore the velocity potential and the stream function 
    all vanish. 
    For the state represented by \eqref{3.e.8}, 
    the complex velocity potential gives 
    \begin{equation}
     W=0, \label{3.e.13}
    \end{equation}
    which expresses {\it fluid at rest} in hydrodynamics. 
    This fluid at rest, however, is not the only one 
    that is physically permissible 
    for a closed state in quantum mechanics, 
    as we can also have flows which are {\it vortical}. 
    For these flows the vorticity may contain singularities 
    in the $xy$-plane. 
    Such flows will be dealt with in Example 3. 

    \paragraph{Example 3. Flows in a central field of force \\}
    As a final example 
    we shall consider the bound state 
    in a certain central field of force. 
    The eigenfunction is, in terms of 
    the polar coordinates $r$, $\theta$, $\varphi$, 
    \begin{equation}
     u_{n l m_l}(r,\theta,\varphi)=
      R_{n l}(r) Y_{l m_l}(\theta,\varphi), \label{3.e.14} 
    \end{equation}
    where the spherical harmonics $Y_{l m_l}(\theta,\varphi)$ 
    are of the form 
    \begin{equation}
     Y_{l m_l}(\theta,\varphi)=C_{l m_l} 
      P_l^{|m_l|}(\cos\theta) e^{i m_l\varphi}, \label{3.e.15}
    \end{equation}
    and $C_{l m_l}$ are the normalizing constants. 
    But $R_{n l}(r)$ for the bound state and 
    the associated Legendre polynomials $P_l^{|m_l|}(\cos\theta)$ 
    are real functions. 
    The polar coordinates $j_r$, $j_\theta$, $j_\varphi$ 
    of \eqref{3.1.2} in a central field of force are thus 
    $$j_r(r,\theta,\varphi)= j_\theta(r,\theta,\varphi)=0,\,\,\, 
    j_\varphi(r,\theta,\varphi)=|u_{n l m_l}(r,\theta,\varphi)|^2 
    \frac{m_l\hslash}{mr\sin\theta}. $$
    In consequence, a simple treatment becomes possible, namely, 
    we may consider the ``velocity'' 
    for a definite direction $\theta$ 
    and then we can introduce the radius $\rho=r\sin\theta$ 
    in above equations 
    and get a problem in two degrees of freedom $\rho$, $\varphi$. 
    The two-dimensional polar coordinates 
    ``$v_\rho$'', ``$v_\varphi$'' of \eqref{3.1.5} give 
    $$\text{``$v_\rho$''}=0,\,\,\, 
    \text{``$v_\varphi$''}=\frac{m_l\hslash}{m\rho}. $$
    The divergence of them readily vanishes. 
    If we transform to two-dimensional polar coordinates 
    $\rho$, $\varphi$, equations \eqref{3.1.12} become 
    \begin{equation}
     \text{``$v_\rho$''}=\frac{1}{\rho} 
      \frac{\partial\Psiv}{\partial\varphi},\,\,\, 
      \text{``$v_\varphi$''}=-\frac{\partial\Psiv}{\partial\rho}, 
      \label{3.e.18} 
    \end{equation}
    and the stream function in a central field of force is thus 
    \begin{equation}
     \Psiv=-\frac{m_l\hslash}{m} \log\rho. \label{3.e.19} 
    \end{equation}
    On the other hand, 
    the vorticity \eqref{3.1.8} satisfies, 
    with the help of \eqref{3.e.18}, 
    \begin{equation}
     \omega=\frac{1}{\rho}\frac{\partial}{\partial\rho}
      \left(\rho\text{``$v_\varphi$''}\right) 
      -\frac{1}{\rho}\frac{\partial}{\partial\varphi}
      \text{``$v_\rho$''} 
      =-\nabla^2 \Psiv, \label{3.e.20} 
    \end{equation}
    where $\nabla^2$ is written for 
    the two-dimensional Laplacian operator 
    $$\nabla^2\equiv\frac{1}{\rho}\frac{\partial}{\partial\rho}
    \left(\rho\frac{\partial}{\partial\rho}\right) 
    +\frac{1}{\rho^2}\frac{\partial^2}{\partial\varphi^2}. $$
    On substituting \eqref{3.e.19} in \eqref{3.e.20} we obtain 
    \begin{equation}
     \omega=\frac{m_l\hslash}{m} \nabla^2\log\rho 
      =2\pi\frac{m_l\hslash}{m} \delta(\vrho), \label{3.e.22}
    \end{equation}
    where $\delta(\vrho)$ is 
    the two-dimensional Dirac $\delta$ function. 
    Thus the vorticity in a central field of force vanishes 
    everywhere except the origin $\rho=0$. 
    This singularity will lie along the quantization axis 
    $\theta=0$ and $\pi$ in three-dimensional space. 
    The velocity potential satisfying \eqref{3.1.9} or \eqref{3.1.10} 
    in a central field of force is 
    \begin{equation}
     \Phiv=\frac{m_l\hslash}{m}\varphi. \label{3.e.23} 
    \end{equation}
    For the state represented by \eqref{3.e.14}, 
    the complex velocity potential \eqref{3.1.13} gives, 
    from \eqref{3.e.23} and \eqref{3.e.19} 
    \begin{align}
     W&=\frac{m_l\hslash}{m}\varphi 
     -i\frac{m_l\hslash}{m}\log\rho \notag\\ 
     &=-i\frac{m_l\hslash}{m}\log z, \label{3.e.24} 
    \end{align}
    since $z=\rho e^{i\varphi}$. 
    According to  hydrodynamics~\cite{batchelor}, 
    this expresses the {\it vortex filament}. 
    The strength of the vortex filament is defined by 
    the {\it circulation} 
    round a closed contour $C$ 
    encircling the singularity at the origin $\rho=0$ 
    \begin{equation}
     \Gammav\equiv\oint_C \text{``$v_\varphi$''} ds. \label{3.e.25} 
    \end{equation}
    We make use of Stokes' theorem, 
    \begin{equation}
     \Gammav=\iint_{\!\!\! S} \omega dS, \label{3.e.26} 
    \end{equation}
    where $S$ is a two-dimensional surface 
    whose boundary is the closed contour $C$. 
    On substituting \eqref{3.e.22} in \eqref{3.e.26} we obtain 
    \begin{equation}
     \Gammav= 2\pi\frac{m_l\hslash}{m}, \label{3.e.27}
    \end{equation}
    where the eigenvalue $m_l$ of a component of 
    the angular momentum is an integer. 
    Equation \eqref{3.e.27} informs us that 
    {\it the circulations are quantized in units of $2\pi\hslash/m$ 
    for the state \eqref{3.e.14} moving in a central field of force}. 
    Equation \eqref{3.e.27} is known as 
    {\it Onsager's Quantization of Circulations}, 
    in superfluidity~\cite{onsager,feynman2}. 
    
    \paragraph{}
    The above examples show the great superiority of 
    the ``velocities'' 
    in dealing with the flows in quantum mechanics. 
    In particular, 
    the two-dimensional quantum flows can be expressed 
    by complex velocity potentials and 
    their analytical properties. 
    Up to the present we have considered 
    only stable systems. 
    For a non-stationary state of the unstable system 
    the probability density \eqref{3.1.1} and 
    the probability current \eqref{3.1.2} are not simple, 
    i.e. they generally depend on the time $t$ and the coordinate $\vr$. 
    However, 
    the work will be simple for such unstable systems, 
    since the ``velocity'' \eqref{3.1.5} does not involve $t$. 
    In fact, 
    for the two-dimensional parabolic potential barrier~\cite{sk4}, 
    as an example of the unstable systems, 
    there is the flow round a right angle 
    that is expressed 
    by the complex velocity potential \eqref{3.e.7} with $n=2$.

\end{document}